\documentclass[letterpaper,titlepage,11pt]{article}
\usepackage{hyperref}
\usepackage{amssymb,amsmath,amsfonts,enumerate}
\usepackage{graphicx}
\usepackage{boxedminipage}

\def\clock{{\count0=\time
           \divide\count0 60
           \ifnum\count0<10 0\fi\the\count0
           \multiply\count0 -60 \advance\count0 \time
           :\ifnum\count0<10 0\fi \the\count0
         }}
\newcommand{\timestamp}{{\small\vbox{\hbox{\tt\jobname.tex}
\hbox{\the\day/\the\month/\the\year, \clock}}}}

\newcommand{\beq}{\begin{equation}}
\newcommand{\eeq}{\end{equation}}
\newcommand{\ben}{\begin{displaymath}}
\newcommand{\een}{\end{displaymath}}
\newcommand{\beqa}{\begin{eqnarray}}
\newcommand{\eeqa}{\end{eqnarray}}
\newcommand{\bea}{\begin{eqnarray}}
\newcommand{\eea}{\end{eqnarray}}
\newcommand{\bean}{\begin{eqnarray*}}
\newcommand{\eean}{\end{eqnarray*}}
\newcommand{\ba}{\begin{array}}
\newcommand{\ea}{\end{array}}
\newcommand{\bi}{\begin{itemize}}
\newcommand{\ei}{\end{itemize}}
\numberwithin{equation}{section}
\begin{document}

\begin{titlepage}
\begin{flushright}
%\timestamp
\end{flushright}
\vskip 2.cm
\begin{center}
{\bf\LARGE{Black Strings Ending }}
\vskip 0.12cm
{\bf\LARGE{on Horizons}} 
\vskip 1.5cm
{\bf Nidal Haddad
}
\vskip 0.5cm
\medskip
\textit{Departament de F{\'\i}sica Fonamental and}\\
\textit{Institut de
Ci\`encies del Cosmos, Universitat de
Barcelona, }\\
\textit{Mart\'{\i} i Franqu\`es 1, E-08028 Barcelona, Spain}\\

\vskip .2 in
\texttt{ nidal@ffn.ub.es}

\end{center}

\vskip 0.3in

\baselineskip 16pt
\date{}

\begin{center} {\bf Abstract} \end{center} 

\vskip 0.2cm 

\noindent We construct an approximate static gravitational solution of the Einstein equations, with negative cosmological constant, describing a test black string stretching from the boundary of the Schwarzschild-AdS$_{5}$ black brane toward the horizon. The construction builds on a derivative expansion method, assuming that the black brane metric changes slowly along the black string direction. We provide a solution up to second order in derivatives and it implies, in particular, that the black string must shrink to zero size at the horizon of the black brane. In the near horizon region of the black brane, where the two horizons intersect, we provide an exact solution of a cone that describes two intersecting horizons at different temperatures. Moreover, we show that this solution equally describes a thin and long black droplet.        

\end{titlepage} \vfill\eject

\setcounter{equation}{0}

\pagestyle{empty}
\small
%\tableofcontents
\normalsize
%\newpage
\pagestyle{plain}
\setcounter{page}{1}

\newpage

\section{Introduction}
General relativity is recovered from string theory at low energies, yet, due to the AdS/CFT correspondence \cite{Maldacena:1998a} its uses go far beyond that. A weakly coupled theory of gravity in AdS spacetime is mapped to a strongly coupled conformal field theory at the boundary of AdS. There are many examples today of theories on AdS spacetime that are mapped to their boundary dual. One example that motivates this paper is the Nambu-Goto string dangling from the boundary of the Schwarzschild-AdS$_{5}$ spacetime toward the horizon \cite{Gubser:2006a,Gubser:2008b,de Boer:2009a}. If one replaces the Nambu-Goto string by a large number of strings, stacked together, then after some limit one expects a configurarion of a black string, with one end at the boundary of Schwarzschild-AdS$_{5}$, dangling toward the black brane horizon. The latter configuration was not found yet. It is a static solution to the Einstein field equations with negative cosmological constant, with the boundary conditions that the region close to the black string and far from the black brane asymptotes to the black string metric in $AdS_{5}$, whereas the region far from the black string asymptotes to the Schwarzschild-AdS$_{5}$ geometry. The configuraion space of black funnels and droplets contains solutions of the former type \cite{Marlof:2010a,Marlof:2010b}. We are going to construct this configuration in a derivative expansion method (similar to \cite{Emparan:2009at,Emparan:2009cs,Bhattacharyya:2008jc}) by assuming that the black brane metric changes slowly along the black string direction (the holographic direction). This is equivalent to say that the black string is a test object in the Schwarzschild-AdS$_{5}$ background. We construct the approximate solution up to second order in derivatives, and we find out, in particular, that the black string must end on the black brane horizon. Furthermore, by providing an exact solution of a self-similar cone that describes an intersection of two horizons at different temperatures $-$ generalizng the cones proposed by Kol \cite{Kol:2004,Kol:2003a,Kol:2003b} $-$ we argue that the black string must end on the black brane in a cone structure. As this configuraion is a natural and a physical one to construct and to study, being one of some immediate generalizations of \cite{Gubser:2006a,Gubser:2008b,de Boer:2009a}, one may think that, although it is a singular solution, this singularity might be smoothed by $\alpha'$ corrections as they become important at the tip of the cone where the curvature radius becomes of the order of the string length. Alternatively, one can view our solution as a critical configuration in the phase space of black funnels and droplets. It can be seen as a thin and long black droplet intersecting the planar black hole in the bulk. In this latter view the singularity is more easily justified, since the cone structure is a necessary ingredient for toplogical phase transitions between black droplets and funnels (when the black droplet and the deformed planar black hole intersect each other at the same temperature and merge) and for mere intersections as well (with no merger thereafter, when the temperatures of the two black objects are different at the intersection) as is the case for our solution; see \cite{Emparan:2011h} for an analogous behavior for a different system.   

We begin in section \ref{sec:basic} by introducing our basic set up and notations. In section \ref{sec:derivative} we introduce our derivatives expansion method, and we obtain, step by step, an approximate solution of a test black string lying along the holographic direction of a Schwarzschild-AdS$_{5}$ black brane up to second order in derivatives. In section \ref{sec:cone} we propose an exact solution of a Ricci-flat cone describing the local geometry near the intersection of two horizons (each at a different temperature). In section \ref{sec:discuss} we discuss the relation of this work to black funnels and droplets and to blackfolds.     
\section{Basic setup and notations}
\label{sec:basic}
Our goal in this work is to construct a static solution to the Einstein equations with negative cosmological constant, in $5$ dimensions, that describes a test black string lying along the radial (holographic) direction of a Schwarzschild-AdS$_{5}$ black brane. The field equations can be written as  
\begin{equation}\label{EE}
E_{\mu\nu}-\frac{6}{R^{2}_{\text{AdS}}}g_{\mu\nu}=0\,,
\end{equation}
where the tensor $E_{\mu\nu}$ is the Einstein tensor $R_{\mu\nu}-\frac{1}{2}Rg_{\mu\nu}$ and $R_{\text{AdS}}$ is the AdS$_{5}$ radius of curvature. The Schwarzschild-AdS$_{5}$ black brane metric is 
\begin{eqnarray}\label{brane1}
G_{\mu\nu}dx^{\mu}dx^{\nu}=\frac{R^{2}_{\text{AdS}}dz^{2}}{z^2f(z)}+\frac{z^2}{R_{\text{AdS}}^{2}}\left[-f(z)dt^2+dr^2+r^{2}d\Omega_{2}^{2}\right]\,,
\end{eqnarray}
with the redshift function given by
\begin{eqnarray}
f(z)=1-\frac{\mu}{z^4}\,.
\end{eqnarray}
The black sting metric in $5-$dimensional flat spacetime is
\begin{eqnarray}
ds^2=dz^{2}-\left(1-\frac{r_{0}}{r}\right)dt^2+\left(1-\frac{r_{0}}{r}\right)^{-1}dr^2+r^{2}d\Omega_{2}^{2}\,,
\end{eqnarray}
and as we are going to see as we advance it will be the seed metric for our construction.
\subsection{Small parameter of the problem}
\label{sec:small}
First of all, we would like to mention at this point that throughout this work we are going to assume that the Schwarzschild radius of the Schwarzschild-AdS$_{5}$ black brane and the curvature radius of AdS$_{5}$ are of the same order of magnitude. Using the mass parameter $\mu$ of the black brane our assumption reads,
\begin{eqnarray}
\mu^{1/4}\sim R_{\text{AdS}}\,.
\end{eqnarray}
Second, we turn to finding the small parameter of our problem. As stated before, we wish to assume that the background metric (\ref{brane1}) changes slowly along the direction of the black string (along the z-axis). It can be easily checked that
\begin{eqnarray}
r^{n}_{0}\partial^{n}_{z} G_{\mu\nu}\propto \frac{r^{n}_{0}}{R^{n}_{\text{AdS}}}\,
\end{eqnarray}
for $\mu\text{,}\nu=t\text{,}z$ and every $n$, and for $\mu\text{,}\nu=r,\Omega$ till $n=2$. For $\mu\text{,}\nu=r,\Omega$ and $n>2$ this quantity vanishes, $r^{n}_{0}\partial^{n}_{z} G_{\mu\nu}=0$. Therefore, if one assumes that $r_{0}\partial_{z} G_{\mu\nu}<<1$ it follows immediately then that 
\begin{eqnarray}
\frac{r_{0}}{R_{\text{AdS}}}<<1\,,
\end{eqnarray}
which we identify as the small dimensionless parameter of the problem. This small parameter really implies that the black string is a test object since its Schwarzschild radius must be much smaller than the curvature radius of the background space, $R_{\text{AdS}}$. Note that in the previous argument we relyed on the fact that the black string in such a configuration can be characterized by its Schwarzschild radius along the $z-$axis. 

\subsection{Field equations}
\label{sec:field}
In what follows we are going to show that the small parameter of the problem brings some simplifications to the field equations if we try to solve them order by order in the small parameter. To see this, take the field equations (\ref{EE}) and make them dimensionless by multiplying both sides by the quantity $r_{0}^2$:
\begin{eqnarray}\label{EE*}
r^{2}_{0}E_{\mu\nu}-6\frac{r^{2}_{0}}{R^{2}_{\text{AdS}}}g_{\mu\nu}=0\,.
\end{eqnarray}
This means that in trying to solve the field equations the cosmological constant term will not enter till second order in the small parameter, or in derivatives equivalently, since it is of order $O(r^{2}_{0}/R^{2}_{\text{AdS}})$. In special, at zeroth and first order in derivatives it will suffice to solve the vacuum Einstein equations
\begin{eqnarray}\label{EE**}
E_{\mu\nu}=0\,.
\end{eqnarray}
The last point makes sense in fact, since the cosmological constant term, being a curvature term, can contribute only when two derivative or more are involved. 
\subsection{Eddington-Finkelstein and adapted coordinates}
\label{sec:EF}
We wish to work throughout with a coordinate system in which the horizon regularity is manifest. For the black string we choose the familiar  Eddington-Finkelstein coordinates, defined through the coordinate transformation $v=t+r+r_{0}\log\left(r/r_{0}-1\right)$,
\begin{eqnarray}
ds^2=dz^{2}-\left(1-\frac{r_{0}}{r}\right)dv^2+2dvdr+r^{2}d\Omega_{2}^{2}\,.
\end{eqnarray}
As for the black brane metric, we should put it, naturally, in a coordinate system \textbf{adapted} to the above black string metric. Performing the coordinate transformation $v=t+r/\sqrt{f(z)}$ on the black brane metric (\ref{brane1}) it takes the following form 
\begin{eqnarray}\nonumber\label{brane2}
ds^2&=&\left[\frac{R^{2}_{\text{AdS}}}{z^2f(z)}-\frac{z^2}{R^{2}_{\text{AdS}}}\left(\frac{r\partial_{z}f(z)}{2f(z)}\right)^{2}\right]dz^{2}-\frac{z^2}{R^{2}_{\text{AdS}}}\left(\frac{r\partial_{z}f(z)}{2f(z)}\right)\left[2\sqrt{f(z)}dvdz-2drdz\right]\\&+&\frac{z^2}{R^{2}_{\text{AdS}}}\left[-f(z)dv^2+2\sqrt{f(z)}dvdr+r^{2}d\Omega_{2}^{2}\right]\,.
\end{eqnarray}
Despite the fact that the latter form of the black brane metric is slightly more complicated than the previous one, (\ref{brane1}), it turns out that it is more convenient to our analysis, as we are going to comment later. 
\section{Derivative expansion method}
\label{sec:derivative}
Look at the following metric that combines both the black string metric and the Schwarzchild-AdS$_{5}$ metric
\begin{eqnarray}\nonumber\label{met}
ds^2&=&\left[\frac{R^{2}_{\text{AdS}}}{z^2f(z)}-\frac{z^2}{R^{2}_{\text{AdS}}}\left(\frac{r\partial_{z}f(z)}{2f(z)}\right)^{2}\right]dz^{2}-\frac{z^2}{R^{2}_{\text{AdS}}}\left(\frac{r\partial_{z}f(z)}{2f(z)}\right)\left[2\sqrt{f(z)}dvdz-2drdz\right]\\&+&\frac{z^2}{R^{2}_{\text{AdS}}}\left[-f(z)\left(1-\frac{r_{0}}{r}\right)dv^2+2\sqrt{f(z)}dvdr+r^{2}d\Omega_{2}^{2}\right]\,.
\end{eqnarray}
This metric does not solve the field equations (\ref{EE}). Nevertheless, it seems to possess all the elements needed to describe the "black string $/$ Schwarzschild-AdS$_{5}$ system", which we are going to list below. Therefore, one hopes that this metric can be corrected somehow so as to make it a solution to the field equations (\ref{EE}). As we are going to show below, applying a derivative expansion method will allow us to correct this metric order by order in derivatives. 

As can be easily checked, the metric (\ref{met}) fulfills the requirement that the black string lies along the $z-$axis of the Schwarzschild-AdS$_{5}$ black brane. If one sets $r_{0}=0$ (or equivalently, if one takes the limit $r>>r_{0}$) the metric reduces to the Schwarzschild-AdS$_{5}$ metric. If instead one sets the mass parameter of the black brane to zero, $\mu=0$, (or equivalently, if one takes the limit $z>>\mu^{1/4}$) the metric reduces to the black string solution in AdS$_{5}$ spacetime. Furthermore, if one focuses (or zooms in) on a slice along the $z-$axis, say the slice $z=z_{c}$ ($z_{c}\neq \mu^{1/4}$), then one recovers the black string solution in flat space. All this suggests, in special, that the above metic respects the correct boundary conditions, the ones appropriate for such a system.

To apply the derivative expansion method promote, first, the Schwarzschild radius of the black string, $r_{0}$, to an unknown function of the $z-$coordinate and then expand it in derivatives around an arbitrary point $z=z_{c}\neq\mu^{1/4}$ as follows,
\begin{eqnarray}
r_{0}=\text{constant}\rightarrow r_{0}(z)=r_{0}(z_{c})+\epsilon \left(z-z_{c}\right)\partial_{z}r_{0}(z_{c})+O(\epsilon^{2})\,.
\end{eqnarray}   
The $\epsilon$ is a formal parameter counting the number of derivatives and it can be set to unity at the end of the calculations. Similarly, expand in derivatives all the other $z$-dependent quantities appearing in the metric,
\begin{eqnarray}
f(z)=f(z_{c})+\epsilon \left(z-z_{c}\right)\partial_{z}f(z_{c})+O(\epsilon^{2})\,,\\
\frac{z^2}{R^{2}_{\text{AdS}}}=\frac{z_{c}^2}{R^{2}_{\text{AdS}}}+\epsilon \left(z-z_{c}\right)\frac{2z_{c}}{R^{2}_{\text{AdS}}}+O(\epsilon^{2})\,.
\end{eqnarray}     
As an explicite illustration, the $zz$ component will read,   
\begin{eqnarray}
g_{zz}\equiv\frac{R^{2}_{\text{AdS}}}{z^2f(z)}-\frac{z^2}{R^{2}_{\text{AdS}}}\left(\frac{r\partial_{z}f(z)}{2f(z)}\right)^{2}=\frac{R^{2}_{\text{AdS}}}{z_{c}^2f(z_{c})}-\epsilon(z-z_{c})\frac{2R^{2}_{\text{AdS}}\left(1+\mu/z_{c}^{4}\right)}{z_{c}^3f(z_{c})^2}+O(\epsilon^{2})\,,
\end{eqnarray} 
where the second term in the $zz$ component does not contribute till second order in derivatives. 
Hence, one can write down the solution to the field equations as 
\begin{eqnarray}\nonumber\label{met1}
ds^2&=&\left[\frac{R^{2}_{\text{AdS}}}{z^2f(z)}-\frac{z^2}{R^{2}_{\text{AdS}}}\left(\frac{r\partial_{z}f(z)}{2f(z)}\right)^{2}\right]dz^{2}-\frac{z^2}{R^{2}_{\text{AdS}}}\left(\frac{r\partial_{z}f(z)}{2f(z)}\right)\left[2\sqrt{f(z)}dvdz-2drdz\right]\\&+&\frac{z^2}{R^{2}_{\text{AdS}}}\left[-f(z)\left(1-\frac{r_{0}(z)}{r}\right)dv^2+2\sqrt{f(z)}dvdr+r^{2}d\Omega_{2}^{2}\right]+...\,,
\end{eqnarray}
where one should understand the above metric as expanded in derivatives and where the dots denote corrections that should be computed order by order in the expansion as we show next.
\subsection{Zeroth order in derivatives}
\label{sec:zeroth}
We are going to show here that at zeroth order the metric (\ref{met1}) solves the field equations (\ref{EE}) with no need for corrections. Indeed, if you assume that the corrections at zeroth order vanish then the metric (\ref{met1}) reads
\begin{eqnarray}\nonumber\label{met0}
ds^2&=&\frac{R^{2}_{\text{AdS}}dz^{2}}{z_{c}^2f(z_{c})}+\frac{z_{c}^2}{R^{2}_{\text{AdS}}}\left[-f(z_{c})\left(1-\frac{r_{0}(z_{c})}{r}\right)dv^2+2\sqrt{f(z_{c})}dvdr+r^{2}d\Omega_{2}^{2}\right]\\
&+&O(\epsilon)\,.
\end{eqnarray}
After a trivial change of units
\beq
z\rightarrow \frac{z_{c}\sqrt{f(z_{c})}}{R_{\text{AdS}}}z\,
\text{,}\,\qquad
v\rightarrow \frac{R_{\text{AdS}}}{z_{c}\sqrt{f(z_{c})}}v\,
\text{,}\,\qquad
r\rightarrow \frac{R_{\text{AdS}}}{z_{c}}r\,
\text{,}\,\qquad
r_{0}\rightarrow \frac{R_{\text{AdS}}}{z_{c}}r_{0}\,
\eeq
it becomes  
\begin{eqnarray}
ds^2=dz^{2}-\left(1-\frac{r_{0}(z_{c})}{r}\right)dv^2+2dvdr+r^{2}d\Omega_{2}^{2}
+O(\epsilon)\,,
\end{eqnarray}
which is the black string metric in flat space, with $R_{\mu\nu}=0$. As was said in section \ref{sec:field}, at zeorth and first order in the derivative expansion one simply needs to solve the vacuum Einstein equations $E_{\mu\nu}=0$, because the cosmological constant term enters at the second order level. That is, at leading order the metric (\ref{met1}) solves the field equations with no need for corrections
\begin{eqnarray}
E_{\mu\nu}-\epsilon^{2}\frac{6}{R^{2}_{\text{AdS}}}g_{\mu\nu}=O(\epsilon)\,.
\end{eqnarray}  
Since the black string metric (which is an exact solution of $E_{\mu\nu}=0$) is the leading order solution in this expansion, it becomes clear that our construction, in principle, consists of  finding corrections to the black string metric order by order in derivatives, so as to describe a black string immersed in the planar black hole background.   
\subsection{First order in derivatives}
\label{sec:first} 
Here one expands the metric (\ref{met1}) up to first order in derivatives, upon which it becomes
\begin{eqnarray}\nonumber\label{met*}
ds^2&=&\left[\frac{R^{2}_{\text{AdS}}}{z^2f(z)}-\frac{z^2}{R^{2}_{\text{AdS}}}\left(\frac{r\partial_{z}f(z)}{2f(z)}\right)^{2}\right]dz^{2}-\frac{z^2}{R^{2}_{\text{AdS}}}\left(\frac{r\partial_{z}f(z)}{2f(z)}\right)\left[2\sqrt{f(z)}dvdz-2drdz\right]\\\nonumber&+&\frac{z^2}{R^{2}_{\text{AdS}}}\left[-f(z)\left(1-\frac{r_{0}(z)}{r}\right)dv^2+2\sqrt{f(z)}dvdr+r^{2}d\Omega_{2}^{2}\right]
+\epsilon h^{(1)}_{\mu\nu}(r)dx^{\mu}dx^{\nu}+O(\epsilon^{2})\\
\,
\end{eqnarray}  
where $h^{(1)}_{\mu\nu}(r)$ are the first-order corrections. The reason that the corrections $h^{(1)}_{\mu\nu}(r)$ are functions of $r$ only, and not of $r$ and $z$, is explained in Appendix[\ref{app:corrections}]. Note that the corrections $h^{(1)}_{\mu\nu}(r)$ are to be solved for so as to make the metric a solution to the field equations and that one should understand the above metric, again, as expanded in derivatives. After choosing the gauge $h^{(1)}_{rr}(r)=h^{(1)}_{\Omega\mu}(r)=0$ (see \cite{Emparan:2010ch}) we found that the field equations are satisfied,
\begin{eqnarray}
E_{\mu\nu}-\epsilon^2\frac{6}{R^{2}_{\text{AdS}}}g_{\mu\nu}=O(\epsilon^2)\,,
\end{eqnarray} 
with the corrections vanishing,
\begin{eqnarray}
h^{(1)}_{\mu\nu}(r)=0\,,
\end{eqnarray}  
and with one constraint on the Schwarzschild radius of the black string,
\begin{eqnarray}
\frac{\partial_{z}r_{0}(z_{c})}{r_{0}(z_{c})}=\frac{1}{2}\frac{\partial_{z}f(z_{c})}{f(z_{c})}\,.
\end{eqnarray} 
This constraint can be integrated to give
\begin{eqnarray}
r_{0}(z)=2M\sqrt{f(z)}\,,
\end{eqnarray}  
where $2M$ is a normalization constant. Note that M is the mass of the black string at the boundary, $z=\infty$. The last constraint implies that the black string must end on the horizon of the Schwarzschild-AdS$_{5}$ black brane, which is one of the main results of this work. An important issue to check in this kind of calculations is the regularity of the black string horizon \cite{Bhattacharyya:2008jc,Emparan:2010ch,Bhattacharyya:2008ji}. By looking at the first order solution found above, one clearly sees that the metric is finite everywhere $-$ remember that we are assuming that $z_{c}>\mu^{1/4}$, so that $f(z_{c})$ is finite . In fact, that is the reason we wanted to work with the E.F coordinates, since they make horizon regularity manifest. 
\subsection{Second order in derivatives}
\label{sec:second} 
At this order the consmological constant term begins to contribute. We continue our previous procedure, and we expand the metric (\ref{met1}) up to second order in derivatives,
\begin{eqnarray}\nonumber\label{met**}
ds^2&=&\left[\frac{R^{2}_{\text{AdS}}}{z^2f(z)}-\frac{z^2}{R^{2}_{\text{AdS}}}\left(\frac{r\partial_{z}f(z)}{2f(z)}\right)^{2}\right]dz^{2}-\frac{z^2}{R^{2}_{\text{AdS}}}\left(\frac{r\partial_{z}f(z)}{2f(z)}\right)\left[2\sqrt{f(z)}dvdz-2drdz\right]\\\nonumber&+&\frac{z^2}{R^{2}_{\text{AdS}}}\left[-f(z)\left(1-\frac{r_{0}(z)}{r}\right)dv^2+2\sqrt{f(z)}dvdr+r^{2}d\Omega_{2}^{2}\right]
+\epsilon^{2} h^{(2)}_{\mu\nu}(r)dx^{\mu}dx^{\nu}+O(\epsilon^{3})\\
\,
\end{eqnarray} 
where $h^{(2)}_{\mu\nu}(r)$ are the second-order corrections. Here, we choose again the gauge $h^{(2)}_{rr}(r)=h^{(2)}_{\Omega\mu}(r)=0$. By solving the field equations (\ref{EE}) up to second order in derivatives,
\begin{eqnarray}
E_{\mu\nu}-\epsilon^2\frac{6}{R^{2}_{\text{AdS}}}g_{\mu\nu}=O(\epsilon^3)\,,
\end{eqnarray}
we have found that the only non-vanishing components are $h^{(2)}_{vv}(r),h^{(2)}_{vr}(r),h^{(2)}_{vz}(r), \text{and } h^{(2)}_{zz}(r)$. We also found that the component $h^{(2)}_{rz}(r)$ is a pure gauge as it does not enter the field equations at this order. After imposing horizon regularity at $r=r_{0}$ $-$ we impose the condition that the corrections $h^{(2)}_{\mu\nu}(r)$ are finite at the horizon $-$ by fixing a certain constant of integration, the solution reads

\begin{eqnarray}\nonumber\label{sol}
&h^{(2)}_{vv}(r)&=
4\frac{\mu r_{0}^2}{R^{6}_{\text{AdS}}}\left[\left(1+\frac{\mu}{2z_{c}^{4}}\right)\frac{r}{r_{0}}+f(z_{c})\left(1-\frac{3r_{0}}{2r}\right)\log\frac{r}{r_{0}}\right]+c_{vv}^{(1)}+\frac{c_{vv}^{(2)}}{r}\,\\
&h^{(2)}_{vr}(r)&=-2\frac{\mu r_{0}^2}{R^{6}_{\text{AdS}}}f(z_{c})^{-\frac{1}{2}}\left[\frac{\mu}{z_{c}^{4}}\frac{r}{r_{0}}+f(z_{c})\log\frac{r}{r_{0}}+\frac{1}{2}f(z_{c})\right]-\frac{c_{vv}^{(1)}}{2\sqrt{f(z_{c})}}\,\\\nonumber
&h^{(2)}_{zz}(r)&=8\frac{\mu}{z_{c}^{4}}\frac{r_{0}^2}{R^{2}_{\text{AdS}}}f(z_{c})^{-1}\left[\frac{r}{r_{0}}+\log\frac{r}{r_{0}}\right]+c_{zz}\,\\\nonumber
&h^{(2)}_{vz}(r)&=c^{(1)}_{vz}+\frac{c^{(2)}_{vz}}{r}\,
\end{eqnarray} 
where the $c_{ab}^{(i)}$'s and $c_{zz}$ are integration constants. In what follows we include some details on how the integration constants are fixed. The constant $c_{zz}$ is a pure gauge because it can be removed by the coordinate change $z\rightarrow z\left(1-\epsilon^{2}\frac{z_{c}^{2}f(z_{c})c_{zz}}{2R^{2}_{\text{AdS}}}\right)$, and so we are free to set it to zero. By performing the coordinate change $v\rightarrow v\left(1+\epsilon^{2}\frac{R^{2}_{\text{AdS}}c_{vv}^{(1)}}{2z_{c}^{2}f(z_{c})}\right)$ one makes the constants $c_{vv}^{(1)}$ and $c_{vv}^{(2)}$ appear only in the $vv$ component in the combination $\frac{c_{vv}^{(2)}+r_{0}c_{vv}^{(1)}}{r}$, which in turn corresponds to a global shift in the temperature of the black string and so we must set it to zero. By the coordinate change $v\rightarrow v-\epsilon^{2}\frac{R^{2}_{\text{AdS}}c^{(2)}_{vz}}{z_{c}^{2}f(z_{c})r_{0}(z_{c})}z$, $z\rightarrow z-\epsilon^{2}\frac{z_{c}^{2}f(z_{c})\left(c^{(1)}_{vz}+c^{(2)}_{vz}/r_{0}(z_{c})\right)}{R^{2}_{\text{AdS}}}v$ and by fixing the pure gauge component $h^{(2)}_{rz}(r)=\frac{c^{(2)}_{vz}}{r_{0}(z_{c})\sqrt{f(z_{c})}}$ one can remove the constants $c_{vz}^{(1)}$ and $c_{vz}^{(2)}$. Altogether, without loss of generality, one can set,
\begin{eqnarray}\label{int}
c_{ab}^{(i)}=c_{zz}=h^{(2)}_{rz}(r)=0\,.
\end{eqnarray}
Moreover, we found that the field equations give a constraint on the quantity $\partial_{z}^{2}r_{0}(z_{c})$, which is  
\begin{eqnarray}
\frac{\partial_{z}^{2}r_{0}(z_{c})}{r_{0}(z_{c})}=\frac{2f\left(z_{c}\right)\partial_{z}^{2}f\left(z_{c}\right)-\left(\partial_{z}f\left(z_{c}\right)\right)^{2}}{4f\left(z_{c}\right)^{2}}\,.
\end{eqnarray}
One can check that the above differential equation is satisfied by the solution found at the first order stage, which we repeat here again,  
\begin{eqnarray}\label{constraint}
r_{0}(z)=2M\sqrt{f(z)}\,.
\end{eqnarray}
\subsection{Properties of the approximate solution}
\label{sec:prop}

Our approximate solution of a test black string ending on the Schwarzschild-AdS$_{5}$ black brane is summarized in Eq(\ref{met**}) and Eqs(\ref{sol}$-$\ref{constraint}). Note that the corrections (\ref{sol}) are multiplyed by an overall factor of $\frac{r_{0}^2}{R^{2}_{\text{AdS}}}$, which is the small parameter squared, as expected at this order. It is worth mentioning, at this stage, two points regarding the validity of our approximation and one point regarding the asymptotics of the solution.
\subsubsection*{Validity of approximation}
\label{sec:val}
\begin{enumerate}    
\item[(i)] 
By looking at the $vv$ component of our solution (after pluging in the constraint $r_{0}=2M\sqrt{f(z)}$), which we write down below in a convenient way,
\begin{eqnarray}\nonumber
g_{vv}=-\frac{z^2f(z)}{R^{2}_{\text{AdS}}}\Biggl[1-\frac{2M\sqrt{f(z)}}{r}
&-&\epsilon^{2}\frac{16\mu }{\sqrt{f(z)}R^{4}_{\text{AdS}}}\frac{M^2}{z^2}\Bigl[\left(1+\frac{\mu}{2z^{4}}\right)\frac{r}{2M}\\ &+&f(z)^{\frac{3}{2}}\left(1-\frac{3r_{0}}{2r}\right)\log\frac{r}{2M\sqrt{f(z)}}\Bigr]\Biggr]\,
\end{eqnarray}  
one can clearly see that the corrections become larger than the leading order term as one approaches the Schwarzschild-AdS$_{5}$ horizon. In other words, since there is an overall factor of $f(z)^{-\frac{1}{2}}$ multiplying the second pair of brackets (representing the corrections), then as $z\rightarrow \mu^{1/4}$ the overall factor diverges. This implies that our approximation breaks down close the black brane horizon. Yet, it does not imply that $z$ must be very far from the horizon for our approximation to be valid - if $z=a\mu^{1/4}$, with $a$ being a constant number of order unity, our approximation will be still valid then. What one can say for sure is that our approximation seems to break down in the near horizon region of the black brane. 

Fortunately, we managed to find, and that is what we are going to show in the next section, an exact solution of a cone describing the local geometry of two intersecting horizons at two different temperatures. By "local geometry" we mean the geometry around the intersection point. That is exactly the region our approximate solution seems to fail to describe. Now, since our approximate solution tells us that the black string horizon must end on the black brane horizon ($r_{0}(z)\rightarrow 0$ as $z\rightarrow\mu^{1/4}$) and since the exact cone solution tells us the same as well, one concludes that the two solutions are two parts (describing two different regions) of one solution. The cone captures the geometry of the near horiozn region of the black brane where the two horizons intersect (it is a Ricci-flat region), while, the approximate solution captures the region outside the near horizon region of the black brane (in especial, the boundary of Schwarzschild-AdS$_{5}$). 
\item[(ii)] 
We must specify what is meant by the asymptotic region $r>>r_{0}$. Since the small parameter of our problem is
\begin{eqnarray}
r_{0}/R_{\text{AdS}} \rightarrow 0\,,
\end{eqnarray}
one should conclude that the maximum value the $r$ coordinate can take is $r\sim R_{\text{AdS}}$. That is, the range of the $r$ coordinate is
\begin{eqnarray}
r\in[r_{0},\sim R_{\text{AdS}}]\,,
\end{eqnarray} 
instead of the more common range, $r\in[r_{0},\infty)$, that one would have encountered had the Schwarzschild-AdS$_{5}$ black brane been absent. 

Hence, in spite of the fact that the corrections (\ref{sol}) seem, to the first sight, to diverge for $r>>r_{0}$, the above clarifications make it clear that  that is not true. One can easily check that as one goes to the correct asymptotic value of $r$ (say  $r\rightarrow R_{\text{AdS}}>>r_{0}$) the corrections remain small and so our approximation remain valid there (Appendix \ref{app:symptotics} shows how this limit is taken).  
\end{enumerate}
\subsubsection*{Asymptotics (thin and long black droplet)}
\label{sec:asym}

Finally, being very important, let us mention some words about the asymptotics of our solution. 

The first asymptotic region of our solution is the boundary of Schwarzschild-AdS$_{5}$ ($z>>R_{\text{AdS}}$). By little work, one can see that if you fix the $r$ coordinate at an arbirary value, and let the coordinate $z$ take larger and larger vales, then at leading order (in $R_{\text{AdS}}/z$ or equivalently in $\mu^{1/4}/z$) the resulting metric is that of a black string in AdS$_{5}$ spacetime. Thus, our solution satisfies the same boundary conditions as black funnels and droplest at the boundary of AdS$_{5}$. (See Appendix \ref{app:symptotics} for details).

The second asymptotic region is the the one with $r\sim R_{\text{AdS}}>>r_{0}$; the region far from the black string. In this region, for generic values of the $z$ coordinate, at leading order in $r_{0}/r\sim r_{0}/R_{\text{AdS}}$, one recovers the Schwarzschild-AdS$_{5}$ geometry. Again, this satisfies the same boundary conditions as black funnels and droplets (see Appendix \ref{app:symptotics} for details).
However, there is a caveat here. The subregion with $r\sim R_{\text{AdS}}>>r_{0}$ and $z\rightarrow \mu^{1/4}$ is tricky as our approximation breaks down for $z\rightarrow \mu^{1/4}$. The point we want to stress in this concern is that there will be no problem with this subregion if you take first the limit $r\sim R_{\text{AdS}}>>r_{0}$, and only afterwards take the $z\rightarrow \mu^{1/4}$ limit.  
  
Hence, by right, one can identify our solution as a thin and long black droplet.   

\section{Exact cone solution}
\label{sec:cone}
We have found in the previous sections, see (\ref{constraint}), that the black string horizon must shrink to zero size at the horizon of the Schwarzschild-AdS$_{5}$ black brane. In spacial, this means that the two horizons intersect at a point\footnote{Note that since we are looking for a local geometry around a point we should be looking for a Ricci-flat geometry.}. If one turns to the Euclidean section and imagines a trajectory in phase space along which the black string approachs the black brane horizon, then one should notice that, as long as the black string did not touch the black brane horizon, the horizon's  $S^{2}$ would be contractible (it shrinks to zero at the tip of the string), while the Euclidean $S^{2}$ would not be contractible (as the two horizon are separated). Once the two horizons meet then the Euclidean $S^{2}$ becomes contractible as well (since it shrinks to zero at the intersection point). All this suggests that we should expect the local geometry to be a cone, the kind proposed in \cite{Kol:2004,Kol:2003a,Kol:2003b}. In $5-$dimensions the cone metric given in \cite{Kol:2004,Kol:2003a,Kol:2003b} can be written (in the Euclidean section) as
\begin{eqnarray}\label{Kol}
ds^2=d\sigma^2+\frac{\sigma^2}{3}\left[d\Omega_{\chi,\tau}^{2}+d\Omega_{\theta,\phi}^{2}\right]
\,,
\end{eqnarray}
which is a cone over $S^{2}\times S^{2}$. However, that is not exactly what we are looking for. The latter cone describes the local geometry of two horizons intersecting at the same temperature, whereas we should be looking for a cone that describes two horizons intersecting at different temperatures; recall that for the black brane the temperature is $T_{\text{Schw-AdS}}=\frac{\mu^{1/4}}{\pi R_{\text{AdS}}^{2}\sqrt{f(z)}}$, while for the black string $T_{\text{BS}}=\frac{1}{8\pi M\sqrt{f(z)}}$ and therefore the temperature of  the latter is much higher than the former.   

In what follows we are going to give an exact Ricci-flat cone solution describing two horizons intersecting at different temperatures, 
\begin{eqnarray}\label{cone}
ds^2=d\sigma^2+\frac{\sigma^2}{L^2}\left[-V(\rho)dt^2+\frac{d\rho^{2}}{V(\rho)}+\rho^{2}d\Omega_{2}^{2}\right]
\,,
\end{eqnarray}
with
\begin{eqnarray}\label{redshift}
V(\rho)=1-\frac{2m}{\rho}-\frac{\rho^2}{L^2}
\,,
\end{eqnarray}
where $L$ and $m$ are arbitrary dimensionful constants. Note that the metric inside the brackets is the Schwarzschild-deSitter metric in $4-$dimensions and so the above metric is a cone over Schwarzschild-dS$_{4}$. In the range of parameters $0<\frac{m^{2}}{L^{2}}<\frac{1}{27}$ the equation
\begin{eqnarray}
V(\rho)=0
\,
\end{eqnarray} 
has two real positive roots , $\rho_{\pm}$ (the inner at the Schwarzschild horizon and the outer at the cosmological one). The temperature of each horizon is given by the formula
\begin{eqnarray}
T_{\pm}=\frac{1}{4\pi}\left|\frac{dV(\rho)}{d\rho}\right|_{\rho_{\pm}}=\frac{1}{2\pi\rho_{\pm}}\left|\frac{m}{\rho_{\pm}}-\frac{\rho_{\pm}^{2}}{L^{2}}\right|\,.
\end{eqnarray}
By fixing $L$ and letting $m$ increase from zero to $\frac{L}{3\sqrt{3}}$, one would be moving in the phase space of solutions from the configuration in which there is no Schwarzschild black hole, but only a cosmological horizon ($\rho_{-}=0$, $\rho_{+}=L$) to the configurarion in which the two horizons overlap ($\rho_{\pm}=L/\sqrt{3}$). The former configuration is infinitesimally close to our situation, because for $m/L<<1$ (corresponding to $r_{0}/R_{\text{AdS}}<<1$) one has that $\rho_{-}=2m\left[1+O(m/L)\right]$ and $\rho_{+}=L\left[1+O(m/L)\right]$,  and consequently the temperatures of the two horizons are, 
\beq
LT_{-}=\frac{L}{8\pi m}\left[1+O(m/L)\right]>>1\,
\text{  and}\,\qquad
LT_{+}=\frac{1}{2\pi}\left[1+O(m/L)\right]=\text{finite}\,
\eeq
which corresponds to our system indeed. In the latter configuration, since the two horizons overlap their temperatures are equal, and so this  is supposed to be identical to the cone given in (\ref{Kol}) as we are going to show next. By taking the Nariai limit of the Schwarzschild-dS$_{4}$ metric appreaing in (\ref{cone}) one gets a cone over the Nariai metric,
\begin{eqnarray}\label{Nariai}
ds^2=d\sigma^2+\frac{\sigma^2}{L^2}\left[-\frac{\rho_{0}^{2}-z^{2}}{\rho_{0}^{2}}dt^2+\frac{\rho_{0}^{2}}{\rho_{0}^{2}-z^{2}}dz^{2}+\rho_{0}^{2}d\Omega_{2}^{2}\right]
\,,
\end{eqnarray}  
where $\rho_{0}\equiv L/\sqrt{3}$ is the value at which the horizons overlap in the $\rho$ coordinate.         
Finally, by performing the coordinate transformation, $z=\rho_{0}\cos\chi$, $t=\rho_{0}\tau$ and going to the Euclidean section one recovers (\ref{Kol}) as promised.
 
\section{Discussion}
\label{sec:discuss}
We have shown that our approximate solution has the same asymptotics as black funnels and droplets. Therefore, we identify our solution as a thin and long black droplet. It is a critical configuration in the phase space of black funnels and droplets. The critical behavior is manifest at the intersection point, where the horizons of the black string and the planar black hole meet, as the geometry there has a cone structure. In contrast to the cones proposed in \cite{Kol:2004,Kol:2003a,Kol:2003b}, which describe the critical point in a topological phase transition from a phase describing two disconnected horizons, to a merged phase with one horizon $-$ the temperatures of the two horizons are equal $-$ in our case as the temperatures of the two horizons are different one expects a mere intersection only (with no merger thereafter). In this view, as we are talking about phase space configurarions, the cone structre is a physical and a universal property (see \cite{Emparan:2011h}), and one would not be worried about the fact that the solution is singular. 
 
It is worth mentioning, in addition, the connection of this work to the blackfolds approach to higher dimensioal black holes \cite{Emparan:2009at,Emparan:2009cs}. Eventhough our approach stems from ideas coming from the blackfolds approach, we did not apply it here. What we did can be considered to be a partial derivation of the blackfolds approach for our specific case. To complete the derivation one would need to compute an effective ($1+1$) stress tensor for the black string which would turn out to be that of a ($1+1$) fluid (in a hydrostatic state) living on a curved background. It would be interseting to do so. In another front, in \cite{Harmark:2010a,Harmark:2010b} the authors applyed the blackfolds approach to setups that share some similarities with ours, yet physically different. In \cite{Harmark:2010a}, for example, the authors studied a static system in which a test black fundamental string has two ends attached to the boundary of Schwarzschild-AdS$_{5}$, by using an effective stress tensor that is obtained from the blackfolds approach.

It would be interesting to generalize the method employed in this work to find approximate bulk solutions for other static systems (when charges and/or other background fields are present for example) and to see if and when the lessons we learned here will change.

\appendix
\section{On the corrections}\label{app:corrections}
In this Appindex we explain the reason (and give a justification) for taking the corrections $h^{(1)}_{\mu\nu}$ in (\ref{met*}) to be functions of $r$ only, and not of $r$ and $z$.  
 
The solution we are looking for is (\ref{met1}), and it can be written, explicitely, in the following form:

\begin{eqnarray}\label{a}
g=g^{(0)}(r,z)+\epsilon g^{(1)}(r,z)+\epsilon^{2} g^{(2)}(r,z)+O(\epsilon^{3})\,,
\end{eqnarray}
where the first term, $g^{(0)}(r,z)$, should be identified with the metric that appears in (\ref{met1}) before the three dots. The rest of the terms, that is, $\epsilon g^{(1)}(r,z)+\epsilon^{2} g^{(2)}(r,z)+O(\epsilon^{3})$, are the corrections and they should be identified with the three dots appearing in (\ref{met1}). Here one sees that the corrections, indeed, depend on both $r$ and $z$, as they should for a general ansatz. However, upon expanding the above metric (\ref{a}) around $z=z_{c}$ it becomes, up to first order in derivatives:
\begin{eqnarray}\label{b}
g=g^{(0)}(r,z_c)+\epsilon \left(z-z_c\right)\partial_{z}g^{(0)}(r,z_c)+\epsilon g^{(1)}(r,z_c)+O(\epsilon^{2})\,,
\end{eqnarray}
and so one notes that, now, the correction term $g^{(1)}(r,z_c)$ is a function of $r$ only, and it is what I denoted in (\ref{met*}) by $h_{\mu\nu}^{(1)}(r)$, namely,
\begin{eqnarray}\label{c}
h_{\mu\nu}^{(1)}(r)\equiv g^{(1)}_{\mu\nu}(r,z_c)\,.
\end{eqnarray} 

This procedure is guaranteed to work to all orders. At order $m$ in derivatives, after expanding the metric (\ref{met1}), or equivalently (\ref{a}), around $z=z_c$ one adds corrections $h_{\mu\nu}^{(m)}$ that are functions of $r$ only, that is,
\begin{eqnarray}
h_{\mu\nu}^{(m)}(r)\equiv g^{(m)}_{\mu\nu}(r,z_c)\,.
\end{eqnarray}

\section{Asymptotics}\label{app:symptotics}
In this section we study the asymptotics of the approximate solution that we have found, given by Eq(\ref{met**}) and Eqs(\ref{sol}$-$\ref{constraint}). There are two asymptotic regions to our solution and we are going to treat them separately below. 

\underline{\textbf{The first asymptotic region}} is the boundary of Schwarzschild-AdS$_{5}$. To take this limit, fix $r$ and take $z$ to infinity
\begin{eqnarray}
z>>\mu^{1/4}\, 
\end{eqnarray}
and remember that in this work we are assuming that the mass of the black brane and the AdS$_{5}$ radius of curvature are of the same order of magnitude,
$\mu^{1/4}\sim R_{\text{AdS}}$. After organizing the solution appropriately, and then assuming $z>>\mu^{1/4}$ one gets
\begin{eqnarray}\nonumber
g_{vv}&=&-\frac{z^2}{R^{2}_{\text{AdS}}}\left[f(z)\left(1-\frac{r_{0}(z)}{r}\right)-
\frac{4\mu }{R^{4}_{\text{AdS}}}\frac{r_{0}^2}{z^2}\left[\left(1+\frac{\mu}{2z^{4}}\right)\frac{r}{r_{0}}+f(z)\left(1-\frac{3r_{0}}{2r}\right)\log\frac{r}{r_{0}}\right]\right]\,\\\nonumber
&=&-\frac{z^2}{R^{2}_{\text{AdS}}}\left[1-\frac{2M}{r}+O\left(\text{max}\left[\mu/z^4,r_{0}^2/z^2\right]\right)\right]\,\\\nonumber
g_{vr}&=&\frac{z^2}{R^{2}_{\text{AdS}}}\left[\sqrt{f(z)}-\frac{2\mu }{\sqrt{f(z)}R^{4}_{\text{AdS}}}\frac{r_{0}^2}{z^2}\left[\frac{\mu}{z^{4}}\frac{r}{r_{0}}+f(z)\log\frac{r}{r_{0}}+\frac{1}{2}f(z)\right]\right]\,\\\nonumber
&=&\frac{z^2}{R^{2}_{\text{AdS}}}\left[1+O\left(\text{max}\left[\mu/z^4,r_{0}^2/z^2\right]\right)\right]\,\\\nonumber
g_{vz}&=&-\frac{z^2}{R^{2}_{\text{AdS}}}\frac{r\partial_{z}f(z)}{2\sqrt{f(z)}}=O\left(r\mu^{\frac{1}{2}}/z^{3}\right)\,\\
g_{rr}&=&0\,\\\nonumber
g_{rz}&=&\frac{z^2}{R^{2}_{\text{AdS}}}\frac{r\partial_{z}f(z)}{2f(z)}=O\left(r\mu^{\frac{1}{2}}/z^{3}\right)\,\\\nonumber
g_{zz}&=&\frac{R^{2}_{\text{AdS}}}{z^2f(z)}-\frac{z^2}{R^{2}_{\text{AdS}}}\left(\frac{r\partial_{z}f(z)}{2f(z)}\right)^{2}+\frac{8\mu}{z^{4}}\frac{r_{0}^2}{f(z)R^{2}_{\text{AdS}}}\left[\frac{r}{r_{0}}+\log\frac{r}{r_{0}}\right]\,\\\nonumber
&=&\frac{R^{2}_{\text{AdS}}}{z^2}+O\left(\text{max}\left[\frac{\mu^{\frac{3}{2}}}{z^6},\frac{r^{2}\mu^{1/2}}{z^{4}},\frac{\mu}{z^4}\frac{r_{0}^{2}}{R^{2}_{\text{AdS}}}\right]\right)\,\\\nonumber
g_{\theta\theta}&=&\frac{z^2}{R^{2}_{\text{AdS}}}r^{2}\,\\\nonumber
g_{\phi\phi}&=&\frac{z^2}{R^{2}_{\text{AdS}}}r^{2}\sin^{2}\theta\,
\end{eqnarray}  
where we have used the fact that $r_{0}=2M\sqrt{f(z)}=2M\left(1+O(\mu/z^4)\right)$ for large z, and where we have set $\epsilon=1$ and replaced everywhere $z_{c}$ by $z$. Remember also that the maximum value $r$ can take is $r\sim R_{\text{AdS}}$. Thus, at leading order one gets the black string metric in AdS$_{5}$
\begin{eqnarray}
ds^2=\frac{R^{2}_{\text{AdS}}}{z^2}dz^{2}+\frac{z^2}{R^{2}_{\text{AdS}}}\left[-\left(1-\frac{2M}{r}\right)dv^2+2dvdr+r^{2}d\Omega_{2}^{2}\right]\,.
\end{eqnarray}
A formal way of taking this limit would be to first perform the following rescalings
\beq
z\rightarrow z/\lambda\,
\text{,}\,\qquad
v\rightarrow \lambda v\,
\text{,}\,\qquad
r\rightarrow \lambda r\,
\text{,}\,\qquad
M\rightarrow \lambda M\,
\eeq
and then to take $\lambda\rightarrow 0$.

\underline{\textbf{The second asymptotic region}} is the one with $r>>r_{0}$, or equivalently, $r\sim R_{\text{AdS}}$, and with a generic value of the $z$ coordinate (i.e. $z>\mu^{1/4}$). After organizing the solution appropriately, and then assuming $r\sim R_{\text{AdS}} $, one gets 
\begin{eqnarray}\nonumber
g_{vv}&=&-\frac{z^2}{R^{2}_{\text{AdS}}}\left[f(z)\left(1-\frac{r_{0}(z)}{r}\right)-
\frac{4\mu }{R^{4}_{\text{AdS}}}\left[\left(1+\frac{\mu}{2z^{4}}\right)\frac{r }{z}\frac{r_{0}}{z}+f(z)\left(1-\frac{3r_{0}}{2r}\right)\frac{r_{0}^2}{z^2}\log\frac{r}{r_{0}}\right]\right]\,\\\nonumber
&=&-\frac{z^2}{R^{2}_{\text{AdS}}}\left[f(z)+O(r_{0}/R_{\text{AdS}})\right]\,\\
g_{vr}&=&\frac{z^2}{R^{2}_{\text{AdS}}}\left[\sqrt{f(z)}-\frac{2\mu }{\sqrt{f(z)}R^{4}_{\text{AdS}}}\left[\frac{\mu}{z^{4}}\frac{r}{z}\frac{r_{0}}{z}+f(z)\frac{r_{0}^2}{z^2}\log\frac{r}{r_{0}}+\frac{1}{2}\frac{r_{0}^2}{z^2}f(z)\right]\right]\,\\\nonumber
&=&\frac{z^2}{R^{2}_{\text{AdS}}}\left[\sqrt{f(z)}+O(r_{0}/R_{\text{AdS}})\right]\,\\\nonumber
g_{zz}&=&\frac{R^{2}_{\text{AdS}}}{z^2f(z)}-\frac{z^2}{R^{2}_{\text{AdS}}}\left(\frac{r\partial_{z}f(z)}{2f(z)}\right)^{2}+\frac{8\mu}{f(z)z^{4}}\left[\frac{r}{R_{\text{AdS}}}\frac{r_{0}}{R_{\text{AdS}}}+\frac{r_{0}^2}{R^{2}_{\text{AdS}}}\log\frac{r}{r_{0}}\right]\,\\\nonumber
&=&\frac{R^{2}_{\text{AdS}}}{z^2f(z)}-\frac{z^2}{R^{2}_{\text{AdS}}}\left(\frac{r\partial_{z}f(z)}{2f(z)}\right)^{2}+O(r_{0}/R_{\text{AdS}})\,
\end{eqnarray}  
where we have used the fact that $r_{0}/z \stackrel{<}{\sim} r_{0}/R_{\text{AdS}}$, which is a direct consequence of the fact that $z\geq\mu^{1/4}$ and $\mu^{1/4}\sim R_{\text{AdS}}$. The rest of the components, being purely background quantities (they do not contain $r_{0}$), remain as they are
\begin{eqnarray}\nonumber
g_{vz}&=&-\frac{z^2}{R^{2}_{\text{AdS}}}\frac{r\partial_{z}f(z)}{2\sqrt{f(z)}}\,\\\nonumber
g_{rr}&=&0\,\\
g_{rz}&=&\frac{z^2}{R^{2}_{\text{AdS}}}\frac{r\partial_{z}f(z)}{2f(z)}\,\\\nonumber
g_{\theta\theta}&=&\frac{z^2}{R^{2}_{\text{AdS}}}r^{2}\,\\\nonumber
g_{\phi\phi}&=&\frac{z^2}{R^{2}_{\text{AdS}}}r^{2}\sin^{2}\theta\,.
\end{eqnarray} 
Altogether, in this region one recovers the background space, the Schwarzschild-AdS$_{5}$ geometry (\ref{brane2}). 

Finally, note that upon taking the above limit we have assumed that $z>\mu^{1/4}$. On the other hand, in the limiting geometry, which is the Schwarzschild-AdS$_{5}$, there is no problem in taking the $z\rightarrow \mu^{1/4}$ limit. Therefore, here there is matter which depends on the order of taking the limits. If you first take the $r>>r_{0}$ limit with $z>\mu^{1/4}$ and later on you take the near horizon limit ($z\rightarrow \mu^{1/4}$) there will be no problem because you will be taking then the near horizon limit of the Schwarzschild-AdS$_{5}$ geometry, which is a well defined limit. If instead, one exchanges the order of taking the limits, a problem will arise since, as was shown in section \ref{sec:val}, our approximation breaks down for $z\rightarrow \mu^{1/4}$.

\end{document}